\documentclass[aps,pra,twocolumn]{revtex4}
\usepackage[english]{babel}
\usepackage[dvips]{graphicx}
\usepackage{enumerate,amsthm,amsmath,amssymb,color,graphicx,bbm,float}
\usepackage{natbib}
\usepackage{hyperref}

\def\d{{\rm d}}

\begin{document}

\title{Finite-size analysis of continuous-variable quantum key distribution}

\author{Anthony Leverrier}
\affiliation{Institut Telecom / Telecom ParisTech,   CNRS LTCI, 46, rue Barrault, 75634 Paris Cedex 13, France}
\affiliation{ICFO-Institut de Cienc\`es Fot\`oniques, 08860 Castelldefels (Barcelona), Spain}

\author{Fr\'ed\'eric Grosshans}
\affiliation{Laboratoire de Photonique Quantique et Mol\'eculaire, ENS de Cachan, UMR CNRS 8735, 94235 Cachan cedex, France}

\author{Philippe Grangier}	
\affiliation{Laboratoire Charles Fabry, Institut d'Optique, CNRS, Univ. Paris-Sud,\\
Campus Polytechnique, RD 128, 91127 Palaiseau Cedex, France}

\date{\today}

\begin{abstract}
The goal of this paper is to extend the framework of finite size analysis recently developed for quantum key distribution to continuous-variable protocols. We do not solve this problem completely here, and we mainly consider the finite size effects on the parameter estimation procedure. Despite the fact that some questions are left open, we are able to give an estimation of the secret key rate for protocols which do not contain a postselection procedure. As expected, these results are significantly more pessimistic than the ones obtained in the asymptotic regime.  However, we show that recent continuous-variable protocols are able to provide fully secure secret keys in the finite size scenario, over distances larger than 50 km.

\end{abstract}
\maketitle

Quantum key distribution (QKD) is a cryptographic primitive that allows two distant parties, Alice and Bob, to generate secret keys despite the presence of a potential eavesdropper \cite{SBC08}.
When, in 1984, Bennett and Brassard invented the first QKD protocol \cite{BB84}, they could only prove that their protocol had to be secure in some appropriate regime (for instance in the unrealistic case where Alice and Bob's data are perfectly correlated) but were not in measure to establish any security proof for a realistic setup at that time. 
During the last 25 years, security proofs have steadily improved and today, \emph{unconditional security}, that is, security guaranteed in an information-theoretical sense has been established for many QKD protocols \cite{SP00,KGR05,RGK05,ren05}, including continuous-variable protocols \cite{RC09}.
A caveat, however, is that any theoretical security proof relies on assumptions, which can be more or less explicit. One such assumption can be for instance that the physical implementation of the protocol behaves as specified, and that no side-channels can be exploited by the adversary. This, unfortunately, is in general impossible to prove \cite{SK09} unless one considers \emph{device-independent} quantum cryptography \cite{ABG07,PAB09}, which is certainly a fascinating theoretical possibility, but highly unpractical.
Another assumption which is often made is that one considers the security of a protocol in the asymptotic regime of infinitely many signals exchanged by Alice and Bob. Here, on the positive side, the general framework to address the problem of finite-size effects already exists. The specific formalism was developed in Renner's PhD thesis \cite{ren05} and subsequently detailed in \cite{SR08} and applied in \cite{CS09}. Note also that another analysis compatible with composable security was developed by Hayashi in the case of BB84 with decoy states \cite{hay07} and later applied to experimental data \cite{HHH07}.
On the negative side, the results of these various papers are quite pessimistic and it is not totally unreasonable to think that the security of all QKD implementations realized until now was in fact jeopardized due to the (way) too short length of the blocks exchanged by Alice and Bob! 

So far, the finite-size analysis has been restricted to discrete-variable protocols, that is, protocols described with a finite dimensional Hilbert space (see Ref. \cite{SS10} for an application to QKD using qudit systems). 

On the other hand, continuous-variable protocols where information is encoded in phase space, appear to be a credible alternative to historical protocols like BB84. Indeed, they are relatively easy to implement, requiring only coherent states generation together with homodyne detection \cite{GVW03} and now display very good performances in terms of achievable distance \cite{LG09,LG10,LG10b}.

The basic idea of continuous-variable QKD is to encode information in phase space. To do so, Alice sends random coherent states to Bob, who decides randomly to measure either one of the quadratures with a homodyne detection. (Note that a variant consists for Bob to perform an heterodyne detection, that is to measure both quadratures simultaneously \cite{WLB04,LG07,SMG07}.) After this measurement, Alice and Bob share classical variables $x$ (the value of the quadrature of the state sent by Alice) and $y$ (Bob's measurement result) from which they can in principle distill a key, once they have exchanged sufficiently many signals. A few protocols of interest have been proved unconditionally secure \cite{RC09}, depending on Alice's modulation which can be either continuous (Gaussian \cite{GVW03} or eight-dimensional \cite{LG10b} in order to allow for a very efficient reconciliation procedure) or discrete (binary \cite{ZHR09} or quaternary \cite{LG09,LG10}). Note that for all such protocols, no postselection of the classical data \cite{SRL02} should be performed, as such a postselection is not compatible with present security proofs \footnote{Remember indeed that protocols with a postselection procedure are only proved to be secure against Gaussian attacks. Unfortunately, in a finite-size context, it is impossible to prove rigorously that an attack is indeed Gaussian. One can likely upper-bound the probability that it is not the case but this is not a trivial task, and it is quite certain that the final secret key rate one could compute would be very small.}.

The goal of the present paper is to give a first finite-size analysis for continuous-variable QKD protocols. In Section \ref{framework}, we rapidly review the framework of \cite{SR08}. In Section \ref{outline}, we describe the outline of a general continuous-variable QKD protocol. Then, we discuss in Section \ref{finite_issues} the main specificities of continuous-variable protocols in the context of finite-size analysis. We proceed in Section \ref{parameter_estimation} with a detailed study of the most important finite-size effect: parameter estimation. Finally, we present the results of this study in Section \ref{finite_results} and give some perspectives.

\section{The general framework for finite size analysis}
\label{framework}

In the following, we note $N$ the total number of signals exchanged by Alice and Bob during the protocol. We note $x$ and $y$ the classical data of Alice and Bob, respectively, after they have measured their quantum states, and $E$ refers to the quantum state of the eavesdropper.

The formalism developed in \cite{ren05} allows for the following generalization of the secret-key rate of a discrete-variable QKD protocol which is secure against collective attacks \cite{SR08}:
\begin{equation}
\label{finite-size-rate}
k = \frac{n}{N} \left( S_{\epsilon_{\text{PE}}}(x|E)  - \frac{\text{leak}_{\text{EC}}}{n} - \Delta(n) \right).
\end{equation}
This key rate has to be compared to the asymptotic key rate $K$ given by
\begin{equation}
K = S(x|E) - H(x|y),
\end{equation}
and four differences can be noticed.

First, only $n$ signals are used for the establishment of the key, out of the $N$ signals exchanged. This is due to the fact that $m = N-n$ signals are used for parameter estimation. This leads to the presence of the prefactor $n/N$ in front of the secret key rate. 
Note that this factor is not very critical: it would become relevant if a QKD protocol were to be implemented and commercialized as it limits the rate of the protocol, but in practice, it has little effect on the final rate, say a factor $1/2$ if $50\%$ of the data are used for parameter estimation. Indeed, the real theoretical challenge today is to decide whether a QKD protocol can be used to distill a secret key for some given conditions (of losses and noise). From this point of view, optimizing every possible parameter in order to maximize the secret key rate seems a little bit premature.

Second, the usual conditional entropy $S(x|E)$ has to be replaced by the expression $S_{\epsilon_{\text{PE}}}(x|E)$ taking into account the finite precision of the parameter estimation. Indeed, whereas the quantum channel can be assumed to be perfectly known in the asymptotic regime, here, this can only be achieved with a finite precision related to the probability $\epsilon_{\text{PE}}$ that the true values of the $n_{\text{PE}}$ channel parameters are not inside the confidence region computed from the parameter estimation procedure. Note there is never unicity of such a confidence region: one is free to optimize his choice among all possible regions compatible with the failure probability $\epsilon_{\text{PE}}$. This choice can be based on the simplicity of the description of the region (for instance the Cartesian product of $n_{\text{PE}}$ intervals corresponding to each estimated parameter), or on an optimization maximizing the final secret key rate, in which case the confidence region is a very general region in the $n_{\text{PE}}$-dimensional space of the parameter space. Unfortunately, such an optimization is often rather complicated to perform, and in general, one chooses confidence regions with very simple shapes (one is referred to Ref. \cite{SMU10} for a discussion of such considerations in the case of BB84). There also exists a trade-off between the desired level of precision of the parameter estimation (in particular, the number $n_{\text{PE}}$ of parameters one considers) and the number $m$ of signals which have to be sacrificed to this end. For instance, it is known that for BB84, in the asymptotic regime, a full tomography of the state increases the secret key rate \cite{WMU08}. This is not necessary true anymore in a finite-size context. 
In \cite{CS09}, the authors suggest that in the limit where $N$ tends to infinity, the optimal number of samples $m$ used for the parameter estimation should be on the order of $\sqrt{N}$. However, for reasonable values of the block length $N$, the number of samples needs to be much larger than $\sqrt{N}$, especially in the case of continuous-variable protocols.

For BB84, only one parameter needs be estimated in theory: the quantum bit error rate (QBER). In practice, however, depending on the implementation, additional parameters might need being estimated especially if weak coherent states are sent instead of true single photons, for instance in the case where the decoy-state technique \cite{LMC05} is applied. Using Hoeffding's inequality for instance, one can find a confidence interval (parameterized by $\epsilon_{\text{PE}}$) for the QBER such that the true value of the parameter is inside the interval with probability at least $1 - \epsilon_{\text{PE}}$. In the case where several parameters must be estimated (for instance in continuous-variable QKD protocols), the notion of confidence interval should be replaced by a multi-dimensional confidence region such that the true value of the parameters lies in the region with a probability at least $1-\epsilon_{\text{PE}}$. Then one needs to compute the minimum value of the conditional entropy $S(x|E)$ compatible with the confidence interval: this gives $S_{\epsilon_{\text{PE}}}(x|E)$. Whereas this procedure is relatively straightforward for the QBER (which is a bounded parameter since $0 \leq \text{QBER} \leq 1$), we will see in the following that the question is more involved for continuous-variable QKD protocols where one needs to estimate \emph{a priori} unbounded parameters such as the excess noise. In this paper, we will consider two parameters to be estimated in continuous-variable QKD: the transmission $T$ and the excess noise $\xi$. In principle, these are not the only parameters to be estimated in a real implementation as one also needs to know Alice's modulation variance (as well as the electronic noise and the quantum efficiency of the detectors if one considers a scenario where Bob's detection is calibrated), but one can reasonably assume that this parameter is relatively well known, in comparison to $T$ and $\xi$.

Third, $\text{leak}_{\text{EC}}$ corresponds to the amount of information which needs to be exchanged by Alice and Bob during the reconciliation phase. This quantity is necessarily equal or larger than the conditional entropy $H(x|y)$, but in practice, it always turns out out be strictly larger than the optimal value. 
We would like to emphasize that the effect of an imperfect reconciliation, which is parameterized by $\text{leak}_{\text{EC}}$ here, was already taken into account for the study of continuous-variable QKD through the so-called \emph{reconciliation efficiency} $\beta$. This reconciliation efficiency gives the fraction of $I(x;y)$, the mutual information between Alice and Bob's data, that the legitimate parties are able to exploit in practice. More precisely, the size of the (identical) bit strings shared by Alice and Bob after the reconciliation procedure is $n \beta I(x;y)$ (the factor $n$ is a reminder that only $n$ couples of data are processed for the key distillation, the rest being used for parameter estimation). This imperfect reconciliation efficiency has been for a long time the cause of the limited range of CV QKD protocols (and still is for the protocols involving a Gaussian modulation). In the case of discrete-variable QKD protocols, $\text{leak}_{\text{EC}}$ is typically modeled as:
\begin{equation}
\text{leak}_{\text{EC}} \approx f_{\text{EC}} H(x|y)  + \frac{1}{n} \log_2(2/\epsilon_{\text{EC}}),
\end{equation} 
where $f_{\text{EC}}>1$ is a parameter characterizing the reconciliation efficiency (in a slightly different way than $\beta$ for continuous-variable reconciliation) and $\epsilon_{\text{EC}}$ is the probability that the reconciliation fails and that this failure goes undetected by Alice and Bob \footnote{In fact, $f_{\text{EC}}$ and $\beta$ are related to each other (for binary variables) through $(f_{\text{EC}}-1) H(x|y) = (1-\beta) I(x;y).$
Using the fact that $I(x;y) = H(x)-H(x|y) = 1-H(x;y)$ for symmetric binary variables, one obtains $f_{\text{EC}} = \frac{1- \beta(1-H(x|y))}{H(x|y)}.$
}. In practice, this probability can be made arbitrarily small.

Finally, the parameter $\Delta(n)$ is related to the security of the privacy amplification. Its value is given by:
\begin{equation}
\label{finite_delta}
\Delta(n) \equiv (2 \text{dim} \, \mathcal{H}_X + 3) \sqrt{\frac{\log_2(2/\bar{\epsilon})}{n}} + \frac{2}{n} \log_2(1/\epsilon_{\text{PA}}),
\end{equation}
where $\mathcal{H}_X$ is the Hilbert space corresponding to the variable $x$ used in the raw key, $\bar{\epsilon}$ is a \emph{smoothing} parameter and $\epsilon_{\text{PA}}$ is the failure probability of the privacy amplification procedure. Both the smoothing parameter $\bar{\epsilon}$  and $\epsilon_{\text{PA}}$ are intermediate parameters which should be optimized numerically. The first term of $\Delta(n)$, that is the square-root term, actually corresponds to the speed of convergence of the smooth min-entropy (which is the appropriate measure of the key length) of an independent and identically distributed (i.i.d.) state (remember that we consider collective attacks here) toward the von Neumann entropy. Indeed, only in the asymptotic limit is the smooth min-entropy of an i.i.d. state equal to its von Neumann entropy. The second term is directly linked to the failure probability $\epsilon_{\text{PA}}$ of the privacy amplification procedure. 

Note that if one was to consider general security, it would be necessary to add to Eq. \ref{finite-size-rate} another correction term linked to the use of the exponential version of de Finetti theorem \cite{ren07}, or to the postselection technique \cite{CKR09} and this would give an even more pessimistic key rate. However, in the case of BB84 for instance, collective attacks are optimal, even in the case of finite size analysis, and such terms are therefore not required. For CV QKD protocols, it is also conjectured but not yet proven that collective attacks are always optimal. Therefore, it makes sense to consider the finite size analysis when the eavesdropper is restricted to collective attacks. Without the proof of the optimality of collective attacks, one could use the bound derived in \cite{RC09} for the application of an exponential version of de Finetti theorem for infinite-dimensional Hilbert spaces. However, because this bound, which is not believed to be tight, leads to very pessimistic results, we prefer not to take it into account here, and therefore limit our analysis to the case of collective attacks.

In the end, one needs to fix an overall security parameter $\epsilon$ for the quantum key distribution protocol. Indeed, contrary to the more usual asymptotic scenario, no such thing as \emph{perfect security} can exist in a finite-size setting, and one is limited to $\epsilon$-security. The (small) parameter $\epsilon$ corresponds to the failure probability of the whole protocol, meaning that the protocol is assured to performed as is supposed to except with a probability at most $\epsilon$. Again, here, we do not consider problems due to an imperfect implementation, which might lead to the existence of side-channels that can be used by an eavesdropper. The failure probability $\epsilon$ can be computed from the various parameters described above, and in the limit of small parameters, one has:
\begin{equation}
\label{finite_epsilon}
\epsilon = \epsilon_{\text{PE}}+\epsilon_{\text{EC}}+\bar{\epsilon} + \epsilon_{\text{PA}}.
\end{equation}

Note that all parameters $\epsilon_{\text{EC}}, \bar{\epsilon}, \epsilon_{\text{PA}}$ and $\epsilon_{\text{PA}}$ can independently be fixed at arbitrarily low values (possibly by increasing the total number $N$ of exchanged signals to a rather high value):
\begin{itemize}
\item $\epsilon_{\text{PE}}$ can be made as low as desired simply by increasing the size of the sample used for parameter estimation (and therefore not used for establishing a key). 
\item $\epsilon_{\text{EC}}$ can be decreased with the following procedure: Alice and Bob simply need to compute a hash of their respective bit strings after the reconciliation and to publicly compare it. This method is very interesting because computing a hash acts like an error amplification: even if the original strings only differ for a few bits out of several millions, their hash will be different with a very high probability. Hence Alice and Bob can check that they share a common bit-string while sacrificing only a negligible quantity of data.
\item $\bar{\epsilon}$ and $\epsilon_{\text{PA}}$ are virtual parameters that can be optimized in the computation. They must simply satisfy both equalities \ref{finite_delta} and \ref{finite_epsilon}.

\end{itemize}
As a consequence, the overall security parameter $\epsilon$ can be chosen arbitrarily small, to a value corresponding to Alice and Bob's wishes. Obviously, this comes at the cost of decreasing the final secret key size.

\section{Outline of a CV QKD protocol in a finite size context}
\label{outline}

Traditionally, in the asymptotic regime, one can make the assumption that the quantum channel is perfectly known, before the transmission is even performed. Hence the optimization of the various free parameters can be made before the exchange of data, and the final secret key rate can be in principle known in advance (as long as the quantum channel behaves as anticipated).

In the finite-size scenario, the situation is quite different. In particular, one does not know in advance the characteristics of the quantum channel. To be fair, even after the exchange of quantum signals, the quantum channel is only partially known: more precisely, a few relevant parameters (QBER for qubit channels, transmission and excess noise for continuous-variable channels) are known to lie inside some confidence regions, except with probability $\epsilon_{\text{PE}}$.

Even if Alice and Bob do not know in advance the properties of the quantum channel they will use, they can guess them with a reasonable accuracy, making the assumption that no eavesdropper will actually try to control the quantum channel. Note that this guess is only used in order to \emph{a priori} optimize various parameters of the protocols and consequently maximize the \emph{expected} secret key rate, under a ``normal use'' of the quantum channel. If an eavesdropper is present, the guess made by Alice and Bob might not be very good, hence leading to an non optimized use of the quantum channel, but the security of the key distribution will not be affected. 

For a continuous-variable QKD protocol, the secret key rate depends (in the asymptotic limit) on three main physical parameters: Alice's modulation variance $V_A$, the transmission of the channel $T$ and the excess noise $\xi$ which corresponds to the noise on Bob's state in excess compared to the shot noise. The idea is to optimize $V_A$ in order to maximize the expected secret key rate. To do that, Alice and Bob can guess the values of $T$ and $\xi$. The transmission can  be evaluated quite precisely with $T \approx \eta 10^{-0.02 d}$ where $\eta$ is the known quantum efficiency of Bob's detection and $d$ is the distance in kilometers between Alice and Bob. Here, we assume an optical fiber with losses of $0.2 \text{dB}$ per kilometer. In practice, Alice and Bob generally have a good estimate of the value of the transmission of the quantum channel before they start the quantum key distribution protocol. The value of $\xi$, on the other hand, depends on the quality of the setup and turns out to be fairly stable from one experimental run of the QKD protocol to the next. Typically, for state-of-the-art implementations, its value is around $1 \%$ of the shot noise \cite{FDD09}. Note that this corresponds to the \emph{expected} value of the excess noise. As we will see later, the measured value might differ significanty from the theoretical value due to a high statistical noise.

At the beginning of the protocol, Alice and Bob agree on a particular value of the overall security parameter $\epsilon$. They also agree on a reconciliation protocol, meaning that they know the parameter $\epsilon_{\text{EC}}$ in advance. Since both $\bar{\epsilon}$ and $\epsilon_{\text{PA}}$ are virtual parameters that have to be optimized afterwards, the rest of the initial (that is before the quantum distribution) optimization consists in studying the parameter $\epsilon_{\text{PE}}$ which quantifies the failure probability of the parameter estimation. This parameter depends on the number $m=N-n$ of samples used for this parameter estimation as well as on some properties of the quantum channels such as the expected true values of the parameters $T$ and $\xi$. Therefore, given the expected behavior of the quantum channel, one can infer the value of $m$ required to obtain a particular value of the parameter $\epsilon_{\text{PE}}$. This in turn puts a lower bound on the block size $N$ necessary to obtain a $\epsilon$-secure secret key. 

At this point, still before the actual start of the quantum key distribution protocol, Alice and Bob can optimize the values of the total number $N$ of signal exchanged, the length of the raw key $n$ as well as the optimal value for Alice's modulation variance $V_A$ in order to maximize the expected secret key rate compatible with the overall security parameter $\epsilon$. The values of $N$, $n$ and $V_A$ can be considered to be fixed at this stage.  

Then, Alice and Bob proceed with the quantum exchange part of the QKD protocol: Alice sends $N$ random coherent states (or $N/2$ for protocols with heterodyne detection) according to the modulation characterizing the protocol (with a modulation variance $V_A$) who measures them with a homodyne (or heterodyne) detection. Bob informs Alice of his measurement choices (that is, for each state, Bob tells Alice whether he measured the $X$ or the $P$ quadrature, or both) and Alice discards the data that Bob did not measure. They publicly disclose $N-n$ of their correlated data in order to estimate the true values of the transmission and excess noise of the quantum channel. They can therefore compute the value of $S_{\epsilon_{\text{PE}}}(y|E)$, the conditional von Neumann entropy of Bob's data (which will be used to form the key in a reverse reconciliation procedure) given Eve's quantum state, which is compatible with the estimated parameters except with probability $\epsilon_{\text{PE}}$. If this value is compatible with a positive secret key rate, Alice and Bob continue with the reconciliation procedure, otherwise they abort the protocol. At the end of the reconciliation procedure, Alice and Bob compute the hash of their respective bit strings (for some well chosen hash function, that is, a randomly chosen hash function from a family such that the equality of both hashes guarantees that the reconciliation procedure worked except with probability $\epsilon_{EC}$ \footnote{We neglect here the (small) number of bits which are revealed trough this procedure.}). If their strings differ (because their hashes differ), Alice and Bob abort the protocol. (Note, however, that Alice and Bob might try to continue with the protocol by exchanging more classical information in order to complete the reconciliation procedure.) If their hashes are identical, Alice and Bob compute the final key size compatible with the security parameter $\epsilon$ (by optimizing over $\bar{\epsilon}$ and $\epsilon_{\text{PA}}$) and perform the privacy amplification, that is, they randomly pick a hash function from a two-universal family of hash functions which outputs a string of length $l$, with $l = k N$ the size of the $\epsilon$-secure secret key that they can extract from their data.

\section{Various issues specific to continuous variables}
\label{finite_issues}

First of all, we would like to emphasize once again that we are only considering here collective attacks and not general attacks. This decision can be motivated by the fact that the optimality of collective attacks has been proven asymptotically and that it is conjectured that this optimality could hold in a finite-size setting. Moreover, the correction terms computed in \cite{RC09} are quite large and likely not tight and might hide interesting effects (such as the dependence of the key rate on parameter estimation) behind purely technical details such as (temporary ?) bounds linked to a particular mathematical proof (exponential version of de Finetti theorem for infinite dimensional Hilbert spaces).

\subsection{Dimensionality}

The main difference between discrete-variable and continuous-variable QKD protocols is obviously the infinite dimensionality of the Hilbert space required to describe CV QKD protocols. This is in general rather problematic when studying the security of CV QKD but it becomes even more annoying when one considers finite size effects. In particular, when one is only interested in asymptotic key rates, the dimension problem can be solved by saying that in the end, everything in the experiment is discrete (even the homodyne detection since the local oscillator has a finite energy). Therefore, one can always theoretically bound the dimension of the relevant Hilbert space by a number large enough and prove that the correction terms due to this large dimension all go to zero in the asymptotic limit. 

Unfortunately, for a finite-size scenario, such an approach 
fails as the convergence toward the asymptotic rate can be heavily slowed for 
high dimension. For instance the finite dimension corresponding to the 
digitalization is often on the order of $d=2^{12}=4096$ if one uses $12$-bit 
analog-to-digital converters. If one compares such a system with a 
two-dimensional one, equation \eqref{finite_delta} tells us that the same 
$\Delta(n)$ is attained for a $n\propto d^2$, \emph{i.e.} for a value of $n$ 
$1.4\cdot10^6$ times higher in the high-dimensional case. The magnitude of such 
factors clearly shows that bounding the dimensionality of the system by a finite
value is not enough for ensuring its security in practical circumstances. 
This dimension should also be as small as 
possible, and it is important to come up with security proofs that are as dimension-independent as possible.

Fortunately, in some continuous variable protocols 
\cite{LAB08-1, LG09,LG10b}, the raw key is encoded on bits, and this allows us
to take  $\text{dim}\, \mathcal{H}_{Y} = 2$ for the numerical evaluation of 
the section \ref{finite_results}.

 When this is not possible, one should be able to replace the real dimension of 
 the system  by its \emph{effective dimension}, but this remains essentially an 
 open question. 
  
Such an effective dimension is generally much smaller than the real dimension and is sufficient to capture the relevant features of the continuous-variable system. Different definitions for this effective dimension can be proposed. Generally however, one defines the effective dimension $d_1^{\text{eff}}$ of a mixed system $\rho$ as:
\begin{equation}
d_1^{\text{eff}}(\rho) \equiv \frac{1}{\text{tr} \rho^2},
\end{equation}
which quantifies the number of states over which $\rho$ is spread \cite{PSW06}. For instance, a totally mixed state over $N$ orthogonal states has $d_1^{\text{eff}} = N$.
A more speculative variant would be to remember that according to \cite{KRS08}, the proper max-entropy $H_{\text{max}}$ of a state is not given by the R\'enyi entropy of order 0 (which diverges for any quantum state of a CV system) but by the R\'enyi entropy of order $1/2$. Therefore one could also define the effective dimension $d_2^{\text{eff}}$ as the exponential of this entropy, that is
\begin{equation}
d_2^{\text{eff}}(\rho) \equiv (\text{tr} \sqrt{\rho})^2.
\end{equation}
The advantage of the first definition is that it relates to the energy of the state, and that it is automatically bounded, for any (physical) state $\rho$. 

In order to be rigorous, the above effective dimension defined above will probably 
be defined under the form of ``smooth dimensions", depending on a small parameter
$\epsilon$, in a way similar to the smooth-entropies introduced in \cite{RW04}, 
which have allowed to look at the finite-size effect. 
In any case, it is intuitively clear that ultimately, it is such effective dimensions that should appear in security proofs of CV QKD protocols. This question certainly needs to be investigated further. We stress again that such a problem, which is rather benign in the asymptotic limit, plays a crucial role in a finite-size analysis.

\subsection{Ill-defined entropies for continuous variables}

Another specificity of CV QKD is that classical entropies are ill-defined for continuous variables: the Shannon entropy has to be replaced by a differential entropy, which is not a practical quantity to analyze secret key rates. For this reason, the expression 
\begin{equation}
S_{\epsilon_{\text{PE}}}(y|E)-\frac{\text{leak}_{\text{EC}}}{n}
\end{equation}
is inadequate for CV QKD. The solution is to rewrite the different quantities in terms of mutual information instead of relative entropies. Hence, the previous expression can be replaced by 
\begin{equation}
\beta I(x:y) - S_{\epsilon_{\text{PE}}}(y:E)
\end{equation}
in the case of a CV protocol. Here  $\beta I(x:y)$ gives the amount of mutual information Alice and Bob were effectively capable to extract through the reconciliation phase: $\beta$ is the reconciliation efficiency which ranges from 0 when no information was extracted to 1 for a perfect reconciliation scheme.  
While $S_{\epsilon_{\text{PE}}}(y|E)$ is defined as the minimum conditional entropy compatible with the statistics given by the parameter estimation except with probability $\epsilon_{\text{PE}}$, $S_{\epsilon_{\text{PE}}}(y:E)$ is naturally defined as the {\em maximum} of the Holevo information compatible with the statistics except with probability $\epsilon_{\text{PE}}$. 
Hence, in the case of a continuous-variable QKD protocol, the secret key rate obtained for a finite size analysis reads:
\begin{equation}
k = \frac{n}{N} \left( \beta I(x:y) - S_{\epsilon_{\text{PE}}}(y:E) - \Delta(n)   \right).
\end{equation}

\subsection{Reconciliation efficiency}

Whereas the question of error-correction was never a crucial issue for DV protocols where it just implied a small correction term, the same is not true for CV protocols \emph{without postselection}. For these schemes, Alice and Bob need to be able to extract their mutual information very efficiently. The absence of efficient reconciliation protocols working in the low Signal-to-Noise Ratio (SNR) regime was, for a long time, the reason why CV protocols could not distribute secret keys as far as their DV counterparts.  To be more precise, a reconciliation protocol is considered efficient if $\beta$ is larger than roughly 80$\%$. For such efficiencies, the correction term appears quite negligible and has a limited impact on the QKD protocol. For a Gaussian modulation, the best known protocols achieve efficiencies higher than 80 $\%$ only for SNR larger than 1 \cite{VCC04,BTM06,LAB08-1}. For lower SNR (relevant to increase the range of the protocol), no good protocol is known for the reconciliation of correlated Gaussian variables. Fortunately, this problem can be solved in this regime by switching to a discrete modulation where efficient protocols are available for all SNR lower than 1 \cite{LG09}.

To summarize, both modulation schemes (Gaussian and discrete) are useful depending on the working distance of the protocol. When working at short distances, a Gaussian modulation should be chosen whereas a discrete modulation is more adapted to reach longer distances. In both cases, the effect of imperfect reconciliation can be taken care of by taking $\beta = 0.8$ which is a conservative value consistent with state of the art reconciliation schemes. 	
Note that one can make the best of both worlds (continuous and discrete modulation) by using a specific eight-dimensional continuous modulation (see Ref. \cite{LG10b} for a detailed presentation of such a protocol).

\section{Parameter estimation}
\label{parameter_estimation}

For discrete-variable QKD protocols, it turns out that the principal finite-size effect, in terms of its consequences on the secret key rate, is the parameter estimation \cite{CS09}. A similar situation is expected for continuous-variable protocols, the main problem being without any doubt, the estimation of the excess noise $\xi$. 

In this section, we study the parameter estimation procedure for continuous-variable protocols, without post\-selection. Quite fortunately, despite being described in an infinite dimensional Hilbert space, these protocols display only a few parameters that need to be estimated: these are the parameters characterizing the covariance matrix of the state shared by Alice and Bob in the entanglement-based version of the protocol \cite{GCW03}. On the positive side, this covariance matrix can be symmetrized through a technique similar to the one explained in \cite{LKG09}, and this symmetrized version is described by only two unknown parameters (in addition to Alice's modulation variance).
On the negative side, in order to be able to estimate the two relevant parameters, it seems that one still has to make the assumption of a Gaussian channel. This does not look as a very constraining assumption as it is known that Alice and Bob can always assume their state to be Gaussian (see \cite{LG09}). However, this result has only been established in the asymptotic limit, and one cannot yet rigorously exclude the (improbable) situation where this result does not hold in general.

A non-Gaussian attack that would exploit such a possible loophole would have to be quite subtle. Indeed, the usual security proof stating that Gaussian states are the ones which minimize the secret key rate cannot be used here because the proof implicitly assumes the knowledge of the covariance matrix. For a \emph{given} covariance matrix, the state maximizing Eve's information is Gaussian. The only possible loophole would be that because the covariance matrix is not perfectly known in a finite-size scenario, there might exist a non-Gaussian state, compatible with the estimated covariance matrix computed with a Gaussian assumption, that would be better for Eve than the Gaussian state estimated by Alice and Bob. 
Let us detail things a little. Let us note $\mathcal{S}_{\epsilon_{\text{PE}}}^g$ the set of states compatible with the results of the parameter estimation, under a Gaussian model, except with probability $\epsilon_{\text{PE}}$. Let us
note $\mathcal{S}_{\epsilon_{\text{PE}}}^{ng}$ the set of states compatible with the results of the parameter estimation, under a general, non-Gaussian model, except with probability $\epsilon_{\text{PE}}$. It is not easy to compare the two sets \emph{a priori}, but one can imagine that they likely get closer and closer as $\epsilon_{\text{PE}}$ goes to 0 (we typically consider $\epsilon_{\text{PE}} = 10^{-10}$). 
In the following, because we make the Gaussian assumption, we consider the Gaussian state $\rho^g \in \mathcal{S}_{\epsilon_{\text{PE}}}^g$ which maximizes Eve's information (note that $\mathcal{S}_{\epsilon_{\text{PE}}}^g$ is not composed of Gaussian states only, but we know that the worst case from Alice and Bob's point of view is Gaussian). However, it is not yet possible to exclude the existence of a non-Gaussian state $\rho^{ng} \in \mathcal{S}_{\epsilon_{\text{PE}}}^{ng}$ such that the secret key rate obtained for $\rho^{ng}$ is strictly lower that the one obtained for $\rho^g$. This is however quite unlikely. 

For this reason, we conjecture that the Gaussian optimality still holds in a non-asymptotic scenario, and in the following, we make the assumption of a Gaussian channel. Again we insist on the point that even if this conjecture were proven wrong, the bounds computed here would still be quite accurate. 

Note, however, that this conjecture was recently established in the case of protocols with a Gaussian modulation \cite{LG09b}.

Our goal, here, is to compute $S_{\epsilon_{\text{PE}}}(y:E)$, the maximal value of the Holevo information between Eve and Bob's classical variable compatible with the statistics except with probability $\epsilon_{\text{PE}}$. The nice property of CV protocols without postselection is that $S(y:E)$ can be bounded from above by a function of two parameters only. More precisely, this function depends on the covariance matrix $\Gamma_{AB}$ of the state $\rho_{AB}$ shared by Alice and Bob in the entanglement-based version of the protocol (see \cite{NGA06,GC06,LG09b} for protocols with a Gaussian modulation and \cite{LG09,LG10} for protocols with a discrete modulation).  One can always assume that $\Gamma_{AB}$ take the following form:
\begin{equation}  
 \Gamma = 
  \begin{pmatrix}
    (V_A+1) \mathbbm{1}_2  & \sqrt{T} Z \sigma_z\\
    \sqrt{T} Z \sigma_z & (T V_A+1 + T \xi) \mathbbm{1}_2\\
  \end{pmatrix},
\end{equation}
where $V_A$ is the variance of Alice's modulation in the {\em prepare and measure} scheme and $T$ and $\xi$ refer to the experimentally estimated effective transmission and excess noise of the channel \cite{LKG09}.  The parameter $Z$ is a function of $V_A$ which depends on the modulation scheme. For instance, one has: $Z_{\text{Gauss}} = \sqrt{V_A^2+2V_A}$ in the case of a Gaussian modulation. For a discrete modulation, $Z$ has a more complicated expression  but turns out to be almost equal to  $Z_{\text{Gauss}}$ for small variances (see \cite{LG09}): for the two-state protocol, one has 
\begin{equation}
Z_2 = V_A \frac{1+e^{-2 V_A}}{\sqrt{1-e^{-2 V_A}}},
\end{equation}
and for the four-state protocol, one has
\begin{equation}
Z_4 = V_A \left( \frac{\lambda_{0}^{3/2}}{\lambda_1^{1/2}} +\frac{\lambda_{1}^{3/2}}{\lambda_2^{1/2}}+\frac{\lambda_{2}^{3/2}}{\lambda_3^{1/2}}+\frac{\lambda_{3}^{3/2}}{\lambda_0^{1/2}} \right),
\end{equation}
where
\begin{equation}
\left\{
\begin{array}{lll}
\lambda_{0,2} &= & \frac{1}{2} e^{-V_A/2}\left( \cosh(V_A/2) \pm \cos(V_A/2) \right) \\
\lambda_{1,3} &= & \frac{1}{2} e^{-V_A/2}\left( \sinh(V_A/2) \pm \sin(V_A/2) \right) .
\end{array}
\right.
\end{equation}
Finally, for the continuous eight-dimensional modulation \cite{LG10b}, one has:
\begin{equation}
Z_8 = \frac{1}{2} e^{-2V_A} \sum_{k=0}^{\infty} \frac{\sqrt{k+4}}{k!}\, V_A^{k+\frac{1}{2}}.
\end{equation}
These various parameters are related through:
\begin{equation}
Z_2 < Z_4 < Z_8 < Z_g.
\end{equation}
In order to compute $S_{\epsilon_{\text{PE}}}(y:E)$, one simply needs to evaluate $\Gamma_{\epsilon_{\text{PE}}}$, the covariance matrix compatible with the data except with probability $\epsilon_{\text{PE}}$ which maximizes the Holevo information between Eve and Bob's classical data.

The estimation of $\Gamma_{\epsilon_{\text{PE}}}$ is made through the sampling of $m \equiv N-n$ couples of correlated variables $(x_i, y_i)_{i=1 \cdots m}$. As we said before, we consider here a normal model for these variables.
Within this model, Alice and Bob's data are linked through the following relation \footnote{the simplicity of this relation comes from the fact that the state $\rho_{AB}$ and consequently the quantum channel are symmetrized.}:
\begin{equation}
y = t x + z
\end{equation}
which is a normal linear model parametrized by $t = \sqrt{T} \in \mathbb{R}$ and where $z$ follows a centered normal distribution with unknown variance $\sigma^2 = 1 + T \xi$. The random variable $x$ can be either a normal random variable with variance $V_A$ in the case of the CV QKD protocol with a Gaussian modulation, or an unbiased Bernoulli random variable taking values $\pm \sqrt{V_A}$ in the case of the two- and four-state protocols (and a similar situation occurs for the continuous eight-dimensional modulation). 
At this point, it is worth considering the dependence of $S(y:E)$ in the variables $t$ and $\sigma^2$. One can in particular check numerically that the following inequalities hold for any value of the modulation variance $V_A$ and for all modulation schemes considered here: 
\begin{equation}
\left.\frac{\partial S(y:E)}{\partial t} \right|_{\sigma^2} < 0 \quad \text{and} \quad \left. \frac{\partial S(y:E)}{\partial \sigma^2}\right|_{t}  > 0.
\end{equation}
This means that one can find the covariance matrix $\Gamma_{\epsilon_{\text{PE}}}$ which minimizes the secret key rate with a probability at least $1 - \epsilon_{\text{PE}}$:
\begin{equation}  
 \Gamma_{\epsilon_{\text{PE}}} = 
  \begin{pmatrix}
    (V_A+1) \mathbbm{1}_2  & t_{\text{min}} Z \sigma_z\\
    t_{\text{min}} Z \sigma_z & (t_{\text{min}}^2 V_A+ \sigma_{\text{max}}^2) \mathbbm{1}_2\\
  \end{pmatrix},
\end{equation}
where $t_{\text{min}}$ and $\sigma_{\text{max}}^2$ correspond respectively to the minimal value of $t$ and the maximal value of $\sigma^2$ compatible with the sampled data, except with probability $\epsilon_{\text{PE}}/2$. Note that this means that the confidence region we consider here is simply a two-dimensional rectangle. One could obviously study more complicated regions that might slightly improve the final key rate. However, here, we prefer to restrict ourselves to this simpler solution which has the advantage of displaying the same features as a more complicated model, but without drowning them under too technical mathematical details. 

Maximum-Likelihood estimators $\hat{t}$ and $\hat{\sigma}^2$ are known for the normal linear model \cite{mon97}:
\begin{equation}
\hat{t} = \frac{\sum_{i=1}^m x_i y_i}{\sum_{i=1}^m x_i^2 }  \quad \text{and} \quad
\hat{\sigma}^2 = \frac{1}{m} \sum_{i=1}^m (y_i-\hat{t} x_i)^2  .
\end{equation}
Moreover, $\hat{t}$ and $\hat{\sigma}^2$ are independent estimators with the following distributions:
\begin{equation}
\label{estimator_pdf}
\hat{t} \sim \mathcal{N}\left(t, \frac{\sigma^2}{\sum_{i=1}^m x_i^2 }\right) \quad \text{and} \quad \frac{m \hat{\sigma}^2}{\sigma^2} \sim \chi^2(m-1),
\end{equation}
where $t$ and $\sigma^2$ are the true values of the parameters. 
This allows us to compute $t_{\text{min}}$, a lower bound for $t$, and $\sigma_{\text{max}}^2$, an upper bound for $\sigma^2$ in the limit of large $m$ \footnote{Indeed, for large $m$, the $\chi^2$ distribution converges to a normal distribution. The approximation is almost exact in our case as we consider values of $m$ (much) larger than $10^6$.}:
\begin{equation}
\label{extremal_values}
\left\{
\begin{array}{lll}
t_{\text{min}} &\approx& \hat{t}- z_{\epsilon_{\text{PE}}/2} \sqrt{\frac{\hat{\sigma}^2}{m \,V_A}}  \\
\sigma_{\text{max}}^2 &\approx& \hat{\sigma}^2 + z_{\epsilon_{\text{PE}}/2} \frac{\hat{\sigma}^2 \sqrt{2}}{\sqrt{m}}  
\end{array}
\right.
\end{equation}
where $z_{\epsilon_{\text{PE}}/2}$ is such that $(1- \mathrm{erf}(z_{\epsilon_{\text{PE}}/2}/\sqrt{2})/2 = \epsilon_{\text{PE}}/2$ and $\mathrm{erf}$ is the \emph{error function} defined as
\begin{equation}
\mathrm{erf}(x) = \frac{2}{\sqrt{\pi}}\int_0^x e^{-t^2} \d t.
\end{equation}

In a given experiment, one can simply compute the values of both estimators $\hat{t}$ and $\hat{\sigma}^2$ and plug them in the Eq. \ref{extremal_values} in order to get the values of $t_{\text{min}}$ and $\sigma_{\text{max}}^2$ and finally the value of $S_{\epsilon_{\text{PE}}}(y:E)$. 
In order to keep analyzing the protocol from a theoretical point of view, we take for $\hat{t}$ and $\hat{\sigma}^2$ their expected values:
\begin{equation}
\begin{array}{lcl}
\mathbb{E}[\hat{t}] &=& \sqrt{T}, \\
\mathbb{E}[\hat{\sigma}^2] &=& 1 + T \xi.
\end{array}
\end{equation}
Using these values, one can compute $t_{\text{min}}$ and $\sigma_{\text{max}}^2$: 
\begin{equation}
\left\{
\begin{array}{lll}
t_{\text{min}} &\approx& \sqrt{T}- z_{\epsilon_{\text{PE}}/2} \sqrt{\frac{1 + T \xi}{m \,V_A}}  \\
\sigma_{\text{max}}^2 &\approx& 1 + T \xi + z_{\epsilon_{\text{PE}}/2} \frac{(1+T\xi) \sqrt{2}}{\sqrt{m}}  .
\end{array}
\right.
\end{equation}

Finally, one gets the covariance matrix $\Gamma_{\epsilon_{\text{PE}}}$ which should be used to compute the expected secret key rate $S_{\epsilon_{\text{PE}}}(y:E)$ in the finite case:
\begin{equation}  
 \mathbb{E}[\Gamma_{\epsilon_{\text{PE}}}] = \Gamma +
  \begin{pmatrix}
    0 & \Delta_Z \sigma_z\\
  \Delta_Z \sigma_z & \Delta_B \mathbbm{1}_2\\
  \end{pmatrix}, 
\end{equation}
with
\begin{equation*}
\left\{
\begin{array}{lll}
\Delta_Z &=& - z_{\epsilon_{\text{PE}}/2} \sqrt{\frac{1 + T \xi}{m \,V_A}} \\
\Delta_B &=& \frac{z_{\epsilon_{\text{PE}}/2}} {\sqrt{m}}((1+T \xi)\sqrt{2} - 2\sqrt{T V_A}) + z_{\epsilon_{\text{PE}}/2}^2 \frac{1+T \xi}{m}  .
\end{array}
\right.
\end{equation*}

At the first order, for long distances, the main effect is clearly the uncertainty on the excess noise. The \emph{effective} excess noise $\Delta_m \xi$ due to the imprecision of the estimation is given by:
\begin{equation}
\label{effective-xi}
\Delta_m \, \xi \approx \frac{z_{\epsilon_{\text{PE}}/2} \sqrt{2}}{T \sqrt{m}}.
\end{equation} 
We will display in the next section the effect of the parameter estimation on the secret key rate, but we can already give a hint about the block length that will generally be required for a given distance. Indeed, from Eq. \ref{effective-xi}, one immediately obtains:
\begin{equation}
m \approx \frac{2  z_{\epsilon_{\text{PE}}/2} ^2}{T^2 \Delta_m \xi^2}.
\end{equation}
For $\epsilon_{\text{PE}} = 10^{-10}$, one has $z_{\epsilon_{\text{PE}}/2} \approx 6.5$, and if one requires $\Delta_m \xi \approx 1/100$, which is a typical value for the true excess noise \cite{FDD09}, then the number of samples required scales as a function of the transmission as
\begin{equation}
m \propto \frac{10^6}{T^2}.
\end{equation}
For instance, if the distance between Alice and Bob is 50 km, then $T=10^{-1}$ which means that one expects the block length to be on the order of $10^8$, which is barely realistic. If the distance is 100 km, then the block length should be on the order of $10^{10}$, which is much more complicated. We will see in Section \ref{finite_results} that the reality is even worse than that.

Before proceeding with the numerical results, we now discuss two technical points linked to the problem of parameter estimation. First, we come back on the assumption made previously that in a typical experiment, the value of the estimators will roughly correspond to the true value of the parameter. Then, we hint on alternative approaches to improve the estimation of the parameters for CV QKD protocols, which might be more economical in terms of the number of samples required, but whose mathematical analysis is much more involved. 

\subsubsection{Expected secret key rate or most probable secret key rate.}

For a given experiment, the secret key rate can be computed and is a function of the observed values of the estimators $\hat{t}$ and $\hat{\sigma}^2$ (but not only). One can write $k_{\text{exp}} = f(\hat{t}, \hat{\sigma}^2)$.
>From a theoretical point of view, that is, without performing the actual experiment, there are two different secret key rates that can be computed.

The first possibility, considered for instance in \cite{CS09}, is to compute the secret key rate $k_1$ obtained for the expected values of the parameters:
\begin{equation}
k_1 \equiv f(\mathbb{E}[\hat{t}], \mathbb{E}[\hat{\sigma}^2]).
\end{equation}
In some sense, this corresponds to what one could call the most probable secret key rate. This interpretation, however, is not correct. 

The correct theoretical secret key rate $k_2$ is given by the expected value of the secret key rate, that is
\begin{equation}
k_2 \equiv \mathbb{E}[f(\hat{t}, \hat{\sigma}^2)].
\end{equation}
Obviously, this value is much more difficult to evaluate in general as one needs to know the probability distributions of both estimators $\hat{t}$ and $\hat{\sigma}^2$, whereas in the case of $k_1$, one just needs to know the expected values. Fortunately, we will see in Section \ref{finite_results} that in fact both values are remarkably close, meaning that one can always safely use $k_1$ as the secret key size.

\subsubsection{More economical parameter estimation procedures.}

An interesting characteristics of continuous-variable QKD protocols is that it might be possible to perform the parameter estimation without sacrificing any data. This is obviously something impossible in discrete-variable QKD where estimating the quantum bit error rate (QBER) requires for Alice and Bob to disclose part of their data. 

In continuous-variable QKD however, the bit used for the raw key is encoded in only a part of Bob's classical data. Let us take the example of the four-state protocol for instance \cite{LG09}. In this case, Bob's data $\{y_i\}_{1\leq i \leq N}$ are real numbers and the raw key elements are simply given by the sign of the variables $y_i$. The absolute value is sent to Alice through the public, authenticated channel, and Alice uses it to perform the reverse reconciliation procedure. 

In the parameter estimation procedure that we described above, Alice and Bob would agree on a certain subset of their data and completely disclose their data in this subset. This means that the absolute values of the rest of Bob's data are not used for this parameter estimation, whereas it manifestly contains information concerning the covariance matrix of the state shared by Alice and Bob. One could certainly use this information to improve the accuracy of the parameter estimation, or equivalently, obtain the same accuracy while using less samples, therefore increasing the final secret key rate. However, the statistics techniques necessary for this study are beyond the scope of this paper, and we do not address this question more extensively here. 

Before concluding this section, we give another possible way to improve upon the parameter procedure presented here. For continuous-variable QKD protocols, it is clear that the critical parameter to estimate is the excess noise. The transmission on the other hand is less critical for two reasons: first, it can be estimated more precisely than the excess noise with the same amount of data (in particular, the relative uncertainty for the transmission is smaller than the one for the excess noise), and second, the secret key rate is much more sensitive to variations in the excess noise than in the transmission. Therefore, one could use the following method in order to estimate the transmission and the parameters:
\begin{itemize}
\item the transmission is estimated through the same procedure as before, with $m$ samples,
\item Bob uses the \emph{totality} of his data to compute an estimation of the variance of his data. Then using the relation $\langle y^2\rangle  = 1 + T V_A + T \xi$ and his estimation of $T$, Bob infers an estimation for the excess noise. 
\end{itemize}
This approach seems better than the one studied above. However, it involves computing two \emph{dependent} estimators and the probability distributions of the estimators (necessary to compute confidence regions) are not known.

Therefore, in this paper, we use a likely suboptimal procedure to perform the parameter estimation, but this procedure allows for the computation of explicit bounds. The question of what is the best parameter estimation procedure in the case of continuous-variable QKD is still open, and is certainly worth investigating further, as it turns out that the parameter estimation is an important problem if one wants to distribute secret keys over long distance, while using realistic block lengths.

\section{Results}
\label{finite_results}

% mathematica file: Finite-size Gaussian_vs_4_state protocol Holevo parano.nb

Before we discuss the results in terms of secret key rate, we start by 
considering the specific influence of $\Delta(n)$, related to privacy 
amplification and defined in eq.~\eqref{finite_delta}.
We then study the influence of the parameter estimation. 

\subsection{Influence of $\Delta(n)$}

\begin{figure}[!ht]
  \centerline{\includegraphics[width=0.95\linewidth]{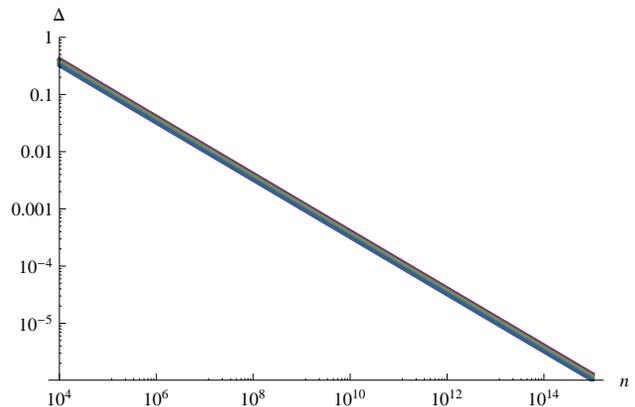}}
	\caption{\label{Delta_finite} (Color online.) Parameter $\Delta(n)$ as a function of $n$ for various values of $\bar{\epsilon}$ and $\epsilon_{\text{PA}}$. From top to bottom, $\bar{\epsilon}=\epsilon_{\text{PA}} = 10^{-6}, 10^{-7},10^{-8},10^{-9}, 10^{-10}$. }
\end{figure}

On Figure \ref{Delta_finite}, we plot the value of the parameter $\Delta(n)$ as a function of $n$, the size of the raw key. Here, we take $\text{dim}\, \mathcal{H}_{Y} = 2$ since for all continuous-variable protocols we consider, the raw key is encoded on bits \cite{LAB08-1, LG09,LG10b}. Among notable features, one sees that the value of $\Delta(n)$
does not critically depend on the parameters $\bar{\epsilon}$ and $\epsilon_{\text{PA}}$ which need to be optimized in theory.
One can easily see from the curve, that for the plotted domain ($n\ge10^4$), 
$\Delta(n)$ is essentially determined by the first term of 
eq.~\eqref{finite_delta}, 
\begin{equation}
	\delta(n)\simeq 7\sqrt{\frac{\log_2(2/\bar{\epsilon})}{n}}
\end{equation}
The important lesson is the large value of $\Delta(n)$, even for (seemingly) quite large sizes of the raw key. In particular, one observes that $\Delta(n)$ is larger than 0.01 for raw key sizes smaller than $10^7$. In practice, this means that if the asymptotic secret key rate is below 0.01 bit per channel use, then if one wants to take into account finite-size effects, one has to use block lengths larger than 10 million in order to be able to claim to have truly distributed a secret key among distance parties. 

\subsection{Influence of the parameter estimation}

\begin{figure}[!ht]
  \centerline{\includegraphics[width=0.95\linewidth]{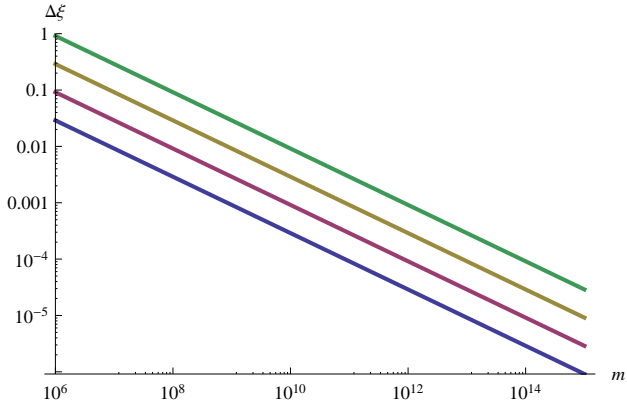}}
	\caption{\label{ksi_finite} (Color online.) Parameter $\Delta_m \xi$ as a function of $m$ for $\epsilon_{\text{PE}}=10^{-10}$ (in fact, $\Delta_m \xi$ does not depend too critically on the precise value of $\epsilon_{\text{PE}}$, and one obtains almost similar plots for $\epsilon_{\text{PE}}=10^{-5}$ for instance). From bottom to top, we consider channel losses of 5 dB, 10 dB, 15 dB and 20 dB. With a perfect homodyne detection (quantum efficiency equal to 1), this is equivalent to distances of 25, 50, 75 and 100 km, respectively.}
\end{figure}

Here, we focus on the value of the \emph{effective} excess noise $\Delta_m \xi$ due to the finite precision of the parameter estimation. On Figure \ref{ksi_finite}, we display this effective excess noise as a function of $m$, the number of samples used in the parameter estimation.
From Figure \ref{ksi_finite}, it is clear that the parameter estimation has a major impact on the final secret key rate. Indeed, the four-state protocol for instance, which can achieve remarkably long distances in the asymptotic limit, requires very low values of the excess noise, typically less than one percent in order to distribute key over distances close to 100 km. Here, we see that such parameters require to sample 10 billion couples of correlated data! For this reason, it is doubtful that continuous-variable QKD is very practical over distances much larger than 100 km, at least with the security proofs presently available. Note that a similar situation is also true for discrete-variable protocols.

\subsection{Secret key rate in the finite-size scenario}

Here, we first consider the secret key rate $k_1$ which is the one obtained if the estimators $\bar{t}$ and $\bar{\sigma}^2$ are equal to their expected values. We do not proceed to a complete optimization of the various parameters since it will not fundamentally change the results and instead take the following values:
\begin{equation}
\left\{
\begin{array}{l}
\epsilon_{\text{EC}} = \bar{\epsilon} = \epsilon_{\text{PA}}  = \epsilon_{\text{PE}} = 10^{-10},\\
\epsilon \approx 10^{-10}, \\
m = n = N/2,\\
\beta = 80, \%\\
\eta = 0.6 .
\end{array}
\right.
\end{equation}
The choice for $\epsilon$ is a very conservative choice, but it turns out that the secret key rate does not depend very critically on $\epsilon$ (a similar observation was made in \cite{SR08}). The choice to use half of the data for the parameter estimation procedure results from the fact that the block size is almost entirely decided by the number of data actually sampled. The reconciliation efficiency of $80 \%$ is a conservative value \cite{LAB08-1,LG09}. Finally, we consider the quantum efficiency of the homodyne detection to be $60 \%$ which corresponds to a typical experimental parameter \cite{FDD09}. 
Moreover, we consider the paranoid mode where the electronic noise is null. Remember that two different models can be considered when discussing the security of CV QKD: either the electronic noise of Bob's detection is considered as regular excess noise (paranoid mode), or it is attributed to Bob's detection (realistic mode). The second case is more natural, but it implies that Bob's detection stage should be calibrated. As a first approximation, the paranoid mode without electronic noise is equivalent (in the asymptotic regime) to the case of a realistic mode where the electronic noise is non negligible but is not supposed to be caused by the action of an eavesdropper. In the finite-size regime, one can still make the assumption that Bob's detection is very well calibrated and that there is virtually no uncertainty on the value of the electronic noise. As a consequence, in order to avoid too many technical details, we present here results obtained in the paranoid scenario without electronic noise. Note that this solution was also chosen in the review by Scarani \emph{et al} \cite{SBC08}.

\begin{figure}[!ht]
	\centerline{\includegraphics[width=0.95\linewidth]{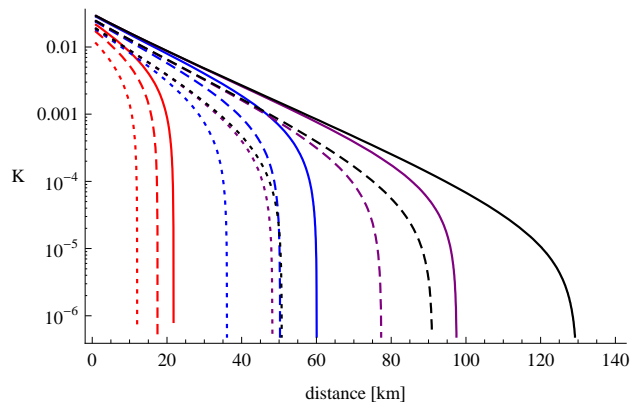}}
	\caption{\label{K4finite} (Color online.) Secret key rate for the four-state protocol. From left to right, the curves correspond respectively to block lengths of $N = 10^8, 10^{10}, 10^{12}$ and $10^{14}$. Full lines, dashed lines and dotted lines correspond respectively to an expected value of the excess noise of 0.001 (optimistic), 0.005 (realistic), 0.01 (conservative). The secret key rate is null for a block length of $10^6$.}
\end{figure}

The secret key rates displayed on Figure \ref{K4finite} correspond to the key rate one can expect if the estimators give the true value of the parameters. As we argued above, a more relevant secret key rate corresponds to the \emph{expected} value computed for the probability distributions of the estimators $\hat{t}$ and $\hat{\sigma}^2$.  

As we already explained, the most important parameter for the final secret key rate is the excess noise. This means that one should mainly consider the probability distribution of the estimator $\hat{\sigma}^2$. 
According to Equation \ref{estimator_pdf}, one has:
\begin{equation}
\frac{m \hat{\sigma}^2}{\sigma^2} \sim \chi^2(m-1).
\end{equation}
For large $m$, the $\chi^{2}$ distribution tends to a normal distribution, which translates into
\begin{equation}
\hat{\sigma}^2 \sim \mathcal{N}\left(\sigma^2, \frac{2 \sigma^4}{m}\right),
\end{equation}
where, as before, $\sigma^2$ corresponds to the true value of the parameter. 
Here, we can therefore compute an approximate value of the expected secret key rate $k_2$ as
\begin{equation}
k_2 = \mathbb{E}[f(\hat{t}, \hat{\sigma}^2)] \approx \mathbb{E}[f(t, \hat{\sigma}^2)] 
\end{equation}
since $\hat{t}$ has a probability distribution peaked around the true value of the parameter $t$ and because $f$ does not depend critically on the value of $t$. 
Using the normality of the random variable $\hat{\sigma}^2$, one obtains
\begin{equation}
k_2 = \int_{-\infty}^{\infty} \frac{1}{2 \sigma^2} \sqrt{\frac{m}{\pi}} \exp \left(- m \frac{(s-\sigma^2)^2}{4 \sigma^4}\right)  f(t, s) \d s.
\end{equation}
% From the relation $\sigma^2 = 1 + T \xi$, one concludes that the observed value of the excess noise $\hat{\xi}$ has the following probability distribution
% \begin{equation}
% \hat{\xi} \sim \mathcal{N}\left(\xi,\frac{2}{T^2 m} \right), 
% \end{equation}
% in the limit where $T \xi \ll 1$. Here, $\xi$ represents the true value of the excess noise (typically between $10^{-3}$ and $10^{-2}$) and $\hat{\xi}$ is the observed value of the excess noise.  

It turns out that the behavior of $k_2$ is numerically indistinguishable from the value of $k_1$. For this reason, we do not display it here. The main consequence is that one can in very good approximation compute the final key rate by considering the expected values of the parameters being estimated.  
This is rather fortunate as computing $k_2$ is much more demanding from a computing point of view than computing $k_1$.

\begin{figure}[!ht]
	\centerline{\includegraphics[width=0.95\linewidth]{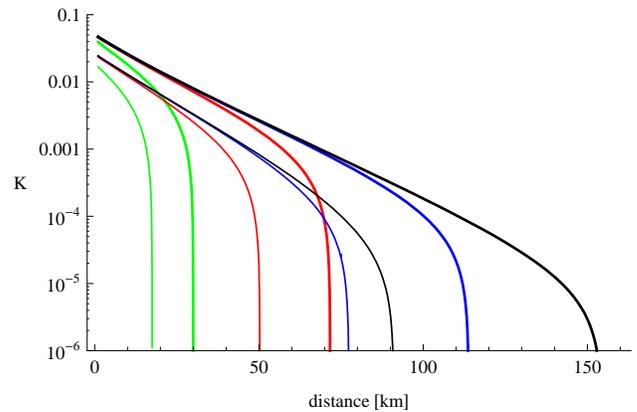}}
	%\centerline{\includegraphics[width=0.95\linewidth]{fig/K4finiteEps10minus5.eps}}
	\caption{\label{K4finite0005bis} (Color online.) Secret key rate for the four-state (thin curves) and eight-dimension (thick curves) protocols obtained for an expected realistic value of the excess noise of 0.005, and for $\epsilon_{\text{PE}} = 10^{-10}$. From left to right, the block length $N$ is equal to $10^{8}, 10^{10}, 10^{12}$ and $10^{14}$. The secret key rate is null for a block length of $10^6$. The eight-dimension protocol is better for all curves. }
\end{figure}

Finally, on Fig. \ref{K4finite0005bis}, we display the results for the recently proposed eight-dimension protocol described in detail in ref. \cite{LG10b}, which outperforms the four-state protocol in realistic cases, and may also be easier to implement. 

\section{Perspectives}

This article  laid down the basis of finite-size analysis of continuous-variable QKD protocols, but several problems remain open, and should be addressed in further studies.

% First, it is not clear that the four-state protocol taken as an example in this article is the best possible one. And actually this is not the case, since an eight-dimension protocol described in ref. \cite{LG10b} performs actually better. For the sake of completeness we have included the performances of this protocol on Fig. 4,  more details are given in \cite{LG10b}. 

First, a basic tool used  in the asymptotic regime is the Gaussian optimality theorem  \cite{GC06,NGA06}, which remain valid for collective attacks, even  in the case of the protocols which do not use a Gaussian modulation, that is, the protocols designed to perform well over long distance \cite{LG09,LG10b}. Whereas using this theorem is completely legitimate in the asymptotic regime \cite{GC06,NGA06}, it is not so clear in a finite-size scenario, where very subtle (and highly improbable) attacks might perform slightly better. Such attacks would have to be based on the idea of fooling Alice and Bob by having them make wrong assumptions in the parameter estimation procedure. This subtle point (given that the blocks have anyway to be very large) deserves further analysis.

Second, and perhaps more importantly, one should prove whether or not collective attacks are optimal in the finite-size setting. If this is the case, then all is for the best, and the bounds derived here are accurate. But if collective attacks are not optimal, then one needs to come up with bounds as tight as possible for coherent attacks. On this subject, it would seem that adapting the postselection technique from \cite{CKR09} to continuous variables (using for instance symmetries in phase space as explained in \cite{LKG09}) would lead to much tighter bounds that the one currently available through an exponential version of de Finetti theorem for infinite dimensional Hilbert spaces \cite{RC09}.

Despite the fact that there are still open problems concerning the finite size analysis of continuous-variable quantum key distribution, some lessons can already be learned. First, it should be emphasized again that  the problem of the reconciliation efficiency at low SNR is essentially solved by new protocols using discrete modulation and high-efficiency reconciliation codes, such as the four-state \cite{LG09} or eight-dimension \cite{LG10b} protocols. Then, as is the case for discrete-variable QKD protocols, the most important remaining  finite-size effect is the limited accuracy of the parameter estimation. For CVQKD, the main issue is  the value of the channel excess noise, which has already a critical effect for values as small as 1\% of shot noise.  Nevertheless, operating windows do exist, provided that both the ``real" excess noise is very small, and the block size is very large. Under such circumstances, secure key distribution over distances larger than 50 km seem quite feasible, especially with the eight-dimension  protocol \cite{LG10b} , as it has been shown on Fig. \ref{K4finite0005bis}.

\section*{Acknowledgments}

The authors acknowledge the hospitality of CQT Singapore during the workshop ``Quantum cryptography with finite resources'' (4-6 December 2008). This work received financial support from Agence Nationale de la Recherche under projects PROSPIQ (ANR-06-NANO-041-05) and SEQURE (ANR-07-SESU-011-01) and from the EU ERC Starting grant PERCENT.

% \bibliography{finite-size}

\end{document}